\documentclass[fleqn,usenatbib]{mnras}

\usepackage{newtxtext,newtxmath}

\usepackage[T1]{fontenc}

\DeclareRobustCommand{\VAN}[3]{#2}
\let\VANthebibliography\thebibliography
\def\thebibliography{\DeclareRobustCommand{\VAN}[3]{##3}\VANthebibliography}

\usepackage{graphicx}	
\usepackage{amsmath}	

\title[VLBI detection of J0240+1952]{VLBI detection of the AE Aqr twin, LAMOST J024048.51+195226.9}

\author[P.-F. Jiang et al.]{Pengfei Jiang,$^{1,2}$
Lang Cui,$^{1,3,4}$\thanks{E-mail: \href{cuilang:cuilang@xao.ac.cn}{cuilang@xao.ac.cn}}
Xiang Liu,$^{1,3,4}$
Bo Zhang,$^{5}$
Yongfeng Huang,$^{6}$
Hongmin Cao,$^{7}$
\newauthor Tao An,$^{5,1}$
Jun Yang,$^{8}$
Fengchun Shu,$^{5}$
Guiping Tan$^{1}$
and Jianping Yuan$^{1}$
\\
$^{1}$Xinjiang Astronomical Observatory, Chinese Academy of Sciences, 150 Science 1-Street, Urumqi 830011, P. R. China\\
$^{2}$School of Astronomy and Space Science, University of Chinese Academy of Sciences, Beijing 100049, P. R. China\\
$^{3}$Key Laboratory of Radio Astronomy, Chinese Academy of Sciences, 150 Science 1-Street, Urumqi 830011, P. R. China\\
$^{4}$Xinjiang Key Laboratory of Radio Astrophysics, 150 Science 1-Street, Urumqi 830011, P. R. China\\
$^{5}$Shanghai Astronomical Observatory, Chinese Academy of Sciences, Shanghai 200030, P. R. China\\
$^{6}$School of Astronomy and Space Science, Nanjing University, Nanjing 210023, P. R. China\\
$^{7}$School of Electronic and Electrical Engineering, Shangqiu Normal University, Wenhua Road 298, Shangqiu, Henan 476000, P. R. China\\
$^{8}$Department of Space, Earth and Environment, Chalmers University of Technology, Onsala Space Observatory, SE-439 92 Onsala, Sweden
}

\date{Accepted 2023 November 17. Received 2023 November 17; in original form 2023 October 21}

\pubyear{2023}

\begin{document}
\label{firstpage}
\pagerange{\pageref{firstpage}--\pageref{lastpage}}
\maketitle

\begin{abstract}
LAMOST J024048.51+195226.9 (J0240+1952) was recently identified as the second AE Aquarii (AE Aqr)-type cataclysmic variable, possessing the fastest known rotating white dwarf. We performed a Very Long Baseline Interferometry (VLBI) observation of J0240+1952 utilizing the European VLBI Network at 1.7\,GHz, to obtain the first view of the radio morphology on mas scale. Our high-resolution VLBI image clearly shows that the radio emission is compact on mas scale ($\lesssim2$\,AU), with no evidence for a radio jet or extended emission. The compact radio source has an average flux density of $\sim0.37$\,mJy, and its brightness temperature is given at $\gtrsim2.3\times10^{7}$\,K, confirming a non-thermal origin. The emission exhibits irregular variations on a time-scale of tens of minutes, similar to the radio flares seen in AE Aqr. The measured VLBI position of J0240+1952 is consistent with that derived from \textit{Gaia}. Our results favour the model in which the radio emission is attributed to a superposition of synchrotron radiation from expanding magnetized blobs of this system.
\end{abstract}




\section{Introduction}

Intermediate polars \citep[e.g.][]{Patterson1994} are a subtype of cataclysmic variable (CV) possessing a white dwarf (WD) with moderate magnetic field strength ($10^{5}$\,G $\lesssim B \lesssim10^{7}$\,G). They are semi-detached binaries in which the WD accretes material from its companion (a lower mass main-sequence star) via Roche lobe overflow \citep{Warner1995}. With the influence of magnetic field, a truncated accretion disc is formed or even a discless accretion \citep{Buckley1995} occurs in these systems. If the moderately strong magnetized WD has a sufficiently fast spin rate, its rapidly rotating magnetosphere may prevent accretion on itself, and the donated material is ejected from this system via a `magnetic propeller' mode \citep{Wynn1995,Eracleous1996,Wynn1997}. However, few magnetic propeller is found in WD systems.

For decades, AE Aquarii (AE Aqr) has been the only confirmed magnetic propeller in CVs \citep{Eracleous1996,Wynn1997}, consisting of a fast rotating WD with a spin period of $\sim33$\,s \citep{Patterson1979}. It is a persistent radio source \citep{Bookbinder1987}, amongst the most radio luminous CVs \citep{Pretorius2021}. Radio flares were found on time-scales of minutes \citep{Bastian1988}. 

Recently, LAMOST J024048.51+195226.9 (J0240+1952) was proposed to be the possible second magnetic CV in a propeller state based on its optical spectra \citep{Thorstensen2020}. The proposal was confirmed by the discovery of high-velocity ($\sim3000$\,km\,s$^{-1}$) emission line flares of this system \citep{Garnavich2021}. A spin period of 24.93\,s was derived for its WD through optical photometry \citep{Pelisoli2022}, implying that J0240+1952 has the fastest rotating WD known to date. A MeerKAT L-band observation detected a bright point-like source on arcsec scale, with a specific radio luminosity of $2.7\times10^{17}$ erg\,s$^{-1}$\,Hz$^{-1}$, higher even than AE Aqr \citep{Pretorius2021}. It was also detected in Karl G. Jansky Very Large Array observations at 2-26\,GHz, showing a persistent radio source varied on a time-scale of a few minutes \citep{Barrett2022}. 

The radio luminous property of these magnetic propellers provides a good opportunity to study them in the radio band. And their radio radiation is typically regarded as non-thermal radiation that arises from synchrotron processes \citep{Bastian1988,Meintjes2003}. However, there are still a variety of debates about the radio emission. \citet{deJager1994} suggested a possible extended radio emission might occur as a relativistic wind is expelled from the polar caps of the magnetic WD. For some accreting WD systems, the detected radio emission has been considered as indirect evidence for a radio jet \citep{Russell2016}. For AE Aqr, Very Long Baseline Interferometry (VLBI) observations found that its radio emission was resolved on a size of orbital radii at times \citep[][and references therein]{deJager1994}, but this has never been confirmed by follow-up studies. Radio emission may have its origin from within the magnetosphere near the WD or the secondary star, with a persistent interaction of the WD's magnetosphere with the secondary star \citep{Bastian1988,Kuijpers1997,Meintjes2003,Meintjes2023}.   

In this letter, we report the first VLBI detection of J0240+1952 with the European VLBI Network (EVN) which allowed us to investigate the morphology and variability of radio emission from this system on mas scale. The observation and data reduction are detailed in Section \ref{sec:section2}. The results are presented and discussed in Section \ref{sec:section3}. Finally, we summarize this study in Section \ref{sec:section4}.

\section{Observation and data reduction}
\label{sec:section2}

A $\sim3$\,h EVN observation of J0240+1952 was conducted at 1.658\,GHz under programme EJ024 on 2021 December 7. Table~\ref{tab:tab1} lists the participating stations and related information. The data were recorded at the rate of 1024\,Mbps (4$\times$32\,MHz subbands, full polarization, two-bit sampling) and correlated in real time \citep[e-VLBI mode,][]{Szomoru2008} by the \textsc{sfxc} software correlator \citep{Keimpema2015} at the Joint Institute for VLBI in Europe. Phase-referencing technique \citep{Beasley1995} was adopted in the observation. J0108+0135 was used as the fringe finder, and J0240+1848 (located 1$\fdg$1 away from J0240+1952, and with a flux density of $\sim0.3$\,Jy) was selected from \textit{Astrogeo Centre}\footnote{\url{http://astrogeo.org/calib/search.html}} as the phase-referencing calibrator. The cycle time is about 3.5\,min, expanding 2.5\,min for J0240+1952 and 1\,min for J0240+1848. Table~\ref{tab:tab2} provides the correlated phase centres for J0240+1952 and its phase-referencing calibrator.

We performed the data reduction by using the Astronomical Image Processing System \citep[\textsc{aips},][]{Greisen2003} and the \textsc{difmap} software \citep{Shepherd1994}. The data severely corrupted by bad weather, instrumental failure, and radio frequency interference were initially flagged. We corrected the ionospheric delay (using Jet Propulsion Laboratory Global Ionospheric Maps) and calibrated the parallactic angle. The visibility amplitude was corrected through antenna gains and system temperatures. Manual instrumental phase and bandpass calibrations were carried out with both J0108+0135 and J0240+1848. We ran a global fringe fitting on J0240+1848 and applied the derived solutions to itself and the target J0240+1952. The calibrated visibility data of J0240+1848 were exported and performed imaging and self-calibrations in \textsc{difmap}. The obtained image of J0240+1848 was then loaded back to \textsc{aips}. We re-ran the global fringe fitting by using the image of J0240+1848 as an input model, correcting the calibrator-structure phase error. The amplitude and phase self-calibrations were also performed with the image of J0240+1848, and the derived solutions were transferred to J0240+1952. Finally, the calibrated visibility data of J0240+1952 were exported and imaged with no self-calibration in \textsc{difmap}.  

\begin{table*}
    \centering
    \caption{Summary of the EVN observation of J0240+1952.}
    \label{tab:tab1}
    \begin{tabular}{llclc}
        \hline
        Project Code& Date         & Frequency (GHz) & Participating Stations$^{a}$               & Duration (h)  \\
        \hline
        EJ024  & 2021 Dec 07 & 1.658  &\texttt{Ef, Jb, Mc, Tr, Ir, Nt, Hh, Wb, O8} & 3       \\[2pt]
        \hline
    \end{tabular}\\
    \footnotesize{$^a$ \texttt{Ef}: Effelsberg (100\,m), \texttt{Jb}: Jodrell Bank MKII (38$\times$25\,m), \texttt{Mc}: Medicina (32\,m), \texttt{Tr}: Torun (32\,m), \texttt{Ir}: Irbene (32\,m), \texttt{Nt}: Noto (32\,m), \texttt{Hh}: Hartebeesthoek (26\,m), \texttt{Wb}: Westerbork (25\,m), \texttt{O8}: Onsala-85 (25\,m).}\\
\end{table*}

\begin{table}
    \centering
    \caption{Source correlation phase centres. $\theta_{\rm d}$ provides the angular distance from J0240+1952.}
    \label{tab:tab2}
    \begin{tabular}{llll}
        \hline
        Source           & $\alpha$(J2000)                  & $\delta$(J2000) & $\theta_{\rm d}$\\
        \hline
        J0240+1952$^{a}$         & $02^{\rm h}40^{\rm m}48\fs529371$ & $+19\degr52\arcmin26\farcs92337$ & ...\\[2pt]
        J0240+1848$^{b}$ & $02^{\rm h}40^{\rm m}42\fs816306$ & $+18\degr48\arcmin00\farcs05508$ & 1$\fdg$1 \\[2pt]
        \hline
    \end{tabular}\\
    \footnotesize{$^{a}$ The position was derived from \citet{Gaia2021}. $^{b}$ The position was taken from \url{http://astrogeo.org/sol/rfc/rfc_2021c/}.}\\
\end{table}

\section{Results and discussion}
\label{sec:section3}

\subsection{Radio emission on mas scale}

\begin{figure}
\centering
\includegraphics[width=0.8\columnwidth]{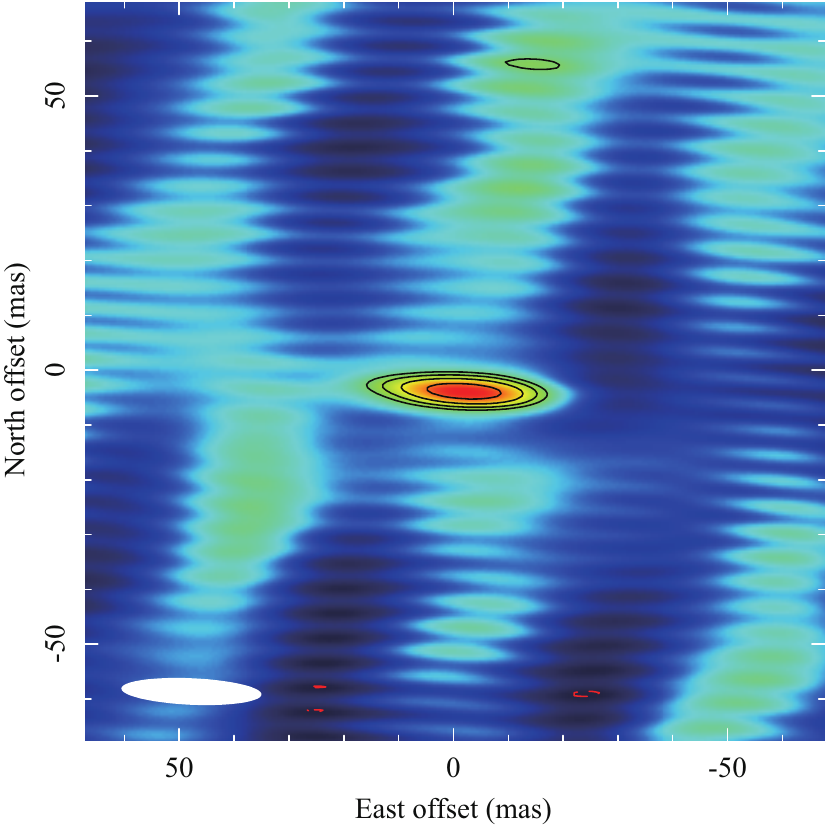}
\caption{Naturally weighted \textsc{clean} map of J0240+1952 obtained with the EVN observation on 2021 December 7 at 1.658\,GHz. The contours start from 3$\sigma$ noise level of 0.1\,mJy\,beam$^{-1}$ and increase by a factor of $2^{1/2}$. The synthesized beam, represented by a white ellipse in the lower left corner, is $25.6 \times 4.8$~mas$^2$, PA = 87$\fdg$5.}
 \label{fig:fig1}
\end{figure}

The naturally weighed EVN image of J0240+1952 is presented in Fig.~\ref{fig:fig1}. 
J0240+1952 is clearly detected as a compact radio source on mas scale, with a signal-to-noise ratio (SNR) of $\sim11$. It shows a point-like structure with a peak flux density, measured in the image plane, of $0.37\pm0.04$\,mJy\,beam$^{-1}$. No extended emission is detected above the 3$\sigma$ noise level of 0.1\,mJy\,beam$^{-1}$. Fitting with a circular Gaussian model to the visibility data gives a source with a total flux density of $\sim0.37$\,mJy and a size of $\sim1.0$\,mas. A specific radio luminosity of $1.7\pm0.2\times10^{17}$ erg\,s$^{-1}$\,Hz$^{-1}$ is derived at the \textit{Gaia} DR3 \citep{Gaia2023} distance of $618\pm27$\,pc, confirming that J0240+1952 is amongst the most radio luminous CVs reported by \citet{Pretorius2021}.

We compared the fitted size to the theoretical minimum resolvable size. The estimated minimum resolvable size $\theta_{\rm lim}$ in a naturally weighted VLBI image is provided by \citet{Lobanov2005}:

\begin{equation}
    \theta_{\rm lim}=\left[ \frac{4\ln{2}}{\pi}\theta_{\rm maj}\theta_{\rm min}\ln{\frac{\rm SNR}{\rm SNR-1}}\right]^{1/2},
    \label{eq:eq1}
\end{equation}
where $\theta_{\rm maj}$ and $\theta_{\rm min}$ are the major and minor axes of the synthesized beam, respectively. For our observation, the fitted size of $\sim1.0$\,mas is smaller than the estimated minimum resolvable size of 3.2\,mas. Thus the radio emission from J0240+1952 is unresolved in this observation. The estimated minimum resolvable size of 3.2\,mas is taken as the upper limit of the source size, and the corresponding radio-emitting region is $\lesssim2$\,AU. We conclude that no evidence is found for a radio jet or extended emission on mas scale.

We calculated the brightness temperature ($T_{\rm B}$) of J0240+1952 as follows \citep[e.g.][]{Bastian1988}: 

\begin{equation}
    T_{\rm B}=9.86\times10^{2}\frac{S_{-3}D^{2}}{r^{2}_{13}\nu^{2}_{9}},
    \label{eq:eq2}
\end{equation}
where $S_{-3}$ is the radio flux density in mJy, $D$ is the distance in units of pc, $r_{13}$ is the radius of the emitting region in units of $10^{13}$\,cm and $\nu_{9}$ is the observing frequency in units of GHz. We derived a brightness temperature of $T_{\rm B}\gtrsim2.3\times10^{7}$\,K, implying a non-thermal origin of the detected radio emission. This supports the idea that the radio emission is non-thermal radiation generated through a synchrotron process \citep[e.g.][]{Bastian1988,Kuijpers1997,Meintjes2003}.

\subsection{Astrometry}

The radio emission of J0240+1952 is detected at the position $\alpha_{\rm J2000} =02^{\rm h}40^{\rm m}48\fs529316\pm1.5$\,mas, $\delta_{\rm J2000}=+19\degr52\arcmin26\farcs91965\pm1.1$\,mas. The coordinate is extracted from the image plane in Fig.~\ref{fig:fig1}. The formal position uncertainty is the combination of the statistical error measured in the image ($\sigma_{\alpha}=1.1$\,mas, $\sigma_{\delta}=0.3$\,mas), the position uncertainty of the phase-referencing calibrator J0240+1848 \citep[$\sigma_{\alpha}=0.13$\,mas, $\sigma_{\delta}=0.14$\,mas;][]{Beasley2002}, the astrometric uncertainty of phase-referencing experiment arising from the station coordinate, Earth orientation and troposphere errors \citep[$\sigma_{\alpha}=0.004$\,mas, $\sigma_{\delta}=0.006$\,mas;][]{Pradel2006} and the astrometric error due to the residual unmodelled ionosphere \citep[$\sim1$\,mas;][]{Rioja2020}.

A full five-parameter astrometric solution of J0240+1952 reported in \textit{Gaia} DR3 \citep{Gaia2023} is $\alpha_{\rm J2016} =02^{\rm h}40^{\rm m}48\fs5310777\pm0.066$\,mas, $\delta_{\rm J2016}=+19\degr52\arcmin26\farcs958763\pm0.057$\,mas, $\pi=1.618\pm0.071$\,mas, $\mu_{\alpha}=-4.060\pm0.097$\,mas\,yr$^{-1}$, and $\mu_{\delta}=-5.969\pm0.081$\,mas\,yr$^{-1}$. We calculate the \textit{Gaia} position to our observing epoch so as to make a direct comparison with the EVN position. The derived \textit{Gaia} position is at $\alpha =02^{\rm h}40^{\rm m}48\fs529310\pm0.58$\,mas, $\delta =+19\degr52\arcmin26\farcs92317\pm0.48$\,mas. We determined that the difference between the EVN position and \textit{Gaia} position is $\Delta\alpha=0.1\pm1.6$\,mas, $\Delta\delta=-3.5\pm1.2$\,mas. 
The difference in right ascension is smaller than $1\sigma$ uncertainty while the difference in declination is within $3\sigma$ uncertainty. We suspect that the difference in declination mainly arises from uncorrected ionospheric effects. As a consequence, the radio and optical positions of J0240+1952 are consistent with each other.

\subsection{Variability}

\begin{figure}
\centering
\includegraphics[width=1\columnwidth]{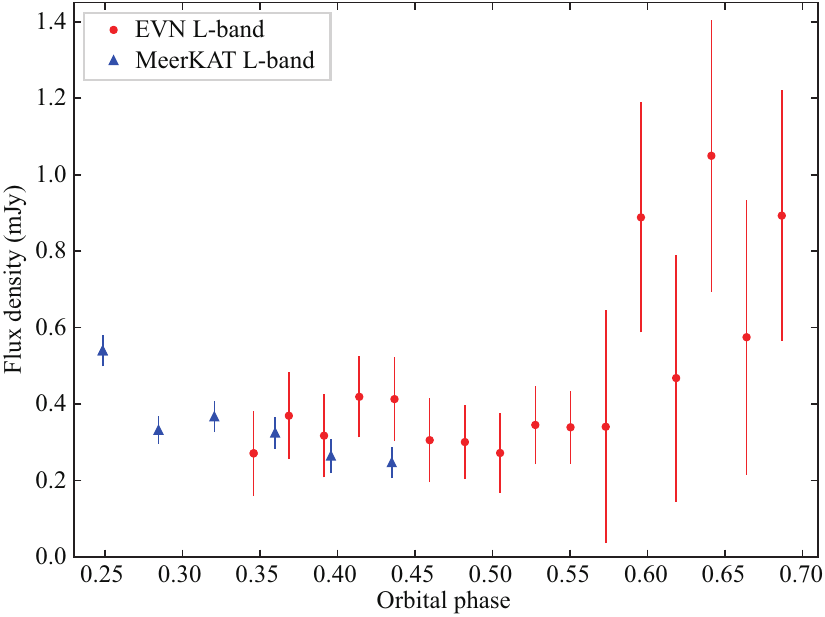}
\caption{Radio light curve of J0240+1952 extracted from the EVN data at 1.7\,GHz and the MeerKAT data at 1.3\,GHz reported by \citet{Pretorius2021}. The integral time of each flux measurement is 10\,min for EVN data and 15\,min for MeerKAT data, respectively.}
 \label{fig:fig2}
\end{figure}

The radio light curve of J0240+1952 derived from EVN data is plotted in Fig.~\ref{fig:fig2}. Each flux measurement has an integral time of 10\,min. The orbital phase is derived by using the ephemeris of \citet{Garnavich2021}, with the observing time converted to Barycentric Julian Date in Barycentric Dynamical Time standard\footnote{\url{http://astroutils.astronomy.osu.edu/time/utc2bjd.html}} \citep{Eastman2010}. Our EVN observation spans orbital phases 0.33-0.71. A loss of sensitivity occurs in the second half of observation due to the absence of some stations (e.g. \texttt{Ef}). 

The radio flux of J0240+1952 varies on a time-scale of tens of minutes (see Fig.~\ref{fig:fig2}). We notice that its flux density increases significantly at the phase of $\sim0.6$. It seems to enter a relatively brighter stage since that, with an average level of $\sim0.8$\,mJy accompanied by obvious fluctuations. Previous MeerKAT L-band observation \citep{Pretorius2021} also found a radio variability on a time-scale of tens of minutes for J0240+1952. \citet{Pretorius2021} proposed that the radio variability could be understood as the end of a radio flare that evolved from optically thick to optically thin with an assumption of synchrotron radiation origin. Based on the same assumption, the rise with rapid fluctuations in intensity in our observation can be explained as a superposition of random flare events caused by synchrotron radiation \citep{Bastian1988}. The radio luminous emission and associated flares observed in J0240+1952 are similar to that found in AE~Aqr, which seems to suggest that these two sources possess comparable magnetic field strengths, probably of $\sim10^6$\,G \citep{Lyutikov2020}. Besides that, the overlapping flare events were also seen in optical photometry, He I emission, and Balmer emission of J0240+1952 \citep{Garnavich2021}. 

The radio flux of J0240+1952 varies randomly at orbital phase in our observation as shown in Fig.~\ref{fig:fig2}. The MeerKAT observation covers orbital phases 0.25-0.44 \citep{Pretorius2021} and has overlapping phases 0.33-0.44 with our EVN observation. There is no correlation of variability between the two observations during the overlapping phases. The flux density decreases in the MeerKAT observation but varies randomly in the EVN observation. We conclude that we see no evidence for orbital modulated radio flux, and it supports the idea that the radio emission should not be confined to the magnetosphere near the WD \citep{Barrett2022}.  

Based on the above discussion, our results favour the model suggested by \citet{Meintjes2003}: the radio flare emission is non-thermal synchrotron radiation generated in a highly magnetized blob-like propeller-ejected outflow of the system. The magnetized blobs originate from the magnetosphere near the secondary star and undergo magnetic interactions with the WD's magnetic field. The magnetized blobs expand and can reach a size of orbital radii. The radio flares come from these expanding magnetized blobs. The variability of the observed radio emission can be explained as a superposition of random flare events from these discrete blobs.

We note that the present radio observations for J0240+1952 only cover a very limited phase range. More observations covering a full period or even several periods should be conducted to reveal whether such variabilities repeat periodically or not, which will be helpful in pinning down the nature of J0240+1952. Moreover, the radio-pulsed nature of AE~Aqr was very recently unveiled by \citet{Meintjes2023}. An underlying pulsed synchrotron emission was detected at the spin period of the WD, apart from the random flare events. If the radio-pulsed emission is a universal feature for magnetic propeller CVs, it could be applicable to J0240+1952. However, J0240+1952 is situated at a considerable distance (618\,pc) and hence has a low flux. Detecting a 24.93\,s periodicity in J0240+1952 is hindered by its low-flux level and the requirement for short integration times. This adds a layer of complexity to our understanding of its radio emission, necessitating future investigations with more sensitive equipment to confirm the potential radio-pulsed emission.

\section{Summary}
\label{sec:section4}

We report the first VLBI detection of the AE Aqr-type CV J0240+1952 using the EVN at 1.658\,GHz. Its radio emission is clearly detected at an average level of $\sim0.37$\,mJy, with an SNR of $\sim11$. High-resolution VLBI image shows that the radio source is unresolved on mas scale, and the corresponding emission region is $\lesssim2$\,AU. There is no evidence that a radio jet or extended emission exists on mas scale. The brightness temperature is $\gtrsim2.3\times10^{7}$\,K, confirming a non-thermal origin of the radio emission. We show that J0240+1952 has irregular variations on a time-scale of tens of minutes, similar to the radio flares seen in AE Aqr. The measured radio position of J0240+1952 agrees with the optical position derived from \textit{Gaia}. Our results suggest that the radio emission can be explained by a superposition of synchrotron radiation arising from expanding magnetized blobs of this system.

We propose that J0240+1952 has a possible underlying radio-pulsed emission, apart from the random radio flares. High-cadence and high-sensitivity monitoring of J0240+1952 is encouraged to verify the existence of these potential radio pulses.

\section*{Acknowledgements}

We thank Zsolt Paragi for helpful suggestions and Petrus Johannes Meintjes for valuable comments.
This work is supported by the National SKA Program of China under grant no. 2022SKA0120102 and the NSFC under grant nos. U2031212 and 12233002. LC acknowledges the support from the CAS `Light of West China' Program (grant no. 2021-XBQNXZ-005). XL acknowledges the support from the National Key R\&D Program of China (grant no. 2023YFE0102300). BZ, YH, and TA acknowledge the support from the Xinjiang Tianchi Program. This work is also partly supported by the Urumqi Nanshan Astronomy and Deep Space Exploration Observation and Research Station of Xinjiang (XJYWZ2303).
The European VLBI Network is a joint facility of independent European, African, Asian, and North American radio astronomy institutes. Scientific results from data presented in this publication are derived from the following EVN project code: EJ024.
e-VLBI research infrastructure in Europe is supported by the European Union’s Seventh Framework Programme (FP7/2007-2013) under grant agreement number RI-261525 NEXPReS. 
This work has made use of data from the European Space Agency (ESA) mission {\it Gaia} (\url{https://www.cosmos.esa.int/gaia}), processed by the {\it Gaia} Data Processing and Analysis Consortium (DPAC, \url{https://www.cosmos.esa.int/web/gaia/dpac/consortium}). Funding for the DPAC has been provided by national institutions, in particular the institutions participating in the {\it Gaia} Multilateral Agreement. 

\section*{Data Availability}

The correlated data of the experiment EJ024 are available in the EVN data archive (\url{http://archive.jive.nl/scripts/portal.php}).
The \textit{Gaia} DR3 data underlying this article are available in the \textit{Gaia} Archive (\url{https://gea.esac.esa.int/archive/}).



\bibliographystyle{mnras}
\bibliography{example} 




\appendix




\bsp	
\label{lastpage}
\end{document}